\documentclass[epj]{svjour}
\usepackage{graphicx}
\begin{document}
\title{Exciton-polariton soliton wavetrains in molecular crystals with dispersive long-range intermolecular interactions}

\author{E. Nji Nde Aboringong \and Alain M. Dikand\'e
}                     
%
%
\institute{Laboratory of Research on Advanced Materials and Nonlinear Sciences (LaRAMaNS),
Department of Physics, \\ Faculty of Science, University of Buea P.O. Box 63 Buea, Cameroon}
\date{Received: date / Revised version: date}
%
\abstract{
The peculiar crystal structure of one-dimensional molecular solids originates from packing of an array of molecules in which intermolecular interactions are dominantly dispersive, including hydrogen-bond, van der Waals and London-type forces. These forces are usually relatively weaker than covalent and ionic bondings such that long-range intermolecular interactions should play important role in dispersion properties of molecular crystals such as polymers and biomolecular chain structures. In this work the effects of long but finite-range intermolecular interactions on single-exciton dispersion energy, and hence on characteristic parameters of periodic soliton trains associated with bound exciton-polariton states in one-dimensional molecular crystals interacting with an electromagnetic field, are investigated. Long-range interactions are shown to quantitatively modify the exciton-polariton soliton amplitudes, width and velocity as a result of shrinkage of the single-exciton energy spectrum. The soliton structures of interest are nonlinear wavetrains, consisting of periodically ordered single-pulse (i.e. bright) or single-kink (i.e. dark) solitons with equal separation between the constituent single-soliton modes. Periodic soliton structures are relevant and best suited for finite-size chain systems, where periodic boundary conditions rule the generation of nonlinear wave profiles. Generally they are of weaker nonlinearity compared to their single-soliton constituents as well established within the framework of their generation via the process of modulational instability. %
} 
\authorrunning{Aboringong et al.} 
\titlerunning{Exciton-polariton soliton wavetrains with long-range interactions}

\maketitle
\section{Introduction}
\label{intro}
Molecular crystals are discrete chains of molecules interacting via van der Waals, hydrogen-bond and London dispersion forces \cite{davyd1,davyd2,davyd3,davyd4}. These molecular solids are characterized by narrow bands with transport properties at the interface between hopping and band transports mediated by excitons \cite{davyd3,davyd4,pout1,pout2}. Excitons in molecular solids are elementary excitations formed from bound electron-hole states that can transport energy without transporting net electric charge, and are known to play important role in free carrier photoexcitations. In fact the transport of energy in form of excitons in molecular crystals is one of the fundamental physical processes characterizing these systems, several related interesting and peculiar properties have indeed been observed in organic molecular chains, biopolymers and protein chains \cite{pout1,pout2,a1,a2,a3,a3p,a4,a4p1,a4p2,a5} as for instance $\alpha$-helix amid chains \cite{edis}. \par
The possible formation of solitons in one-dimensional ($1D$) molecular crystals with multi-exciton scatterings \cite{2ea,2eb,2ec} has been discussed in the past, taking into account the Pauli character of exciton operators \cite{stoy1}. It has been shown that the inclusion of exciton-exciton dimer term in the exciton Hamiltonian with intramolecular vibrations and nearest-neighbour interactions, leads to two possible nonlinear structures namely bright and dark exciton solitons. As emphasized in ref. \cite{stoy1}, the existence and stability of any of this particular type of soliton is determined by specific conditions on values of the system parameters including soliton velocity, and to this last point it was established that neither kind of soliton (i.e. bright or dark) can exist if the short-range intermolecular interaction is repulsive. \par
The interaction of excitons with light has also been extensively discussed in the past \cite{kham1,kham2}, considering the process as a mean to regulate certain functions of molecular crystals as well as to enhance excitonic transport in these materials. In biological systems \cite{stoy2,stoy3,a7} in particular frequency selective effects of light on protein activation, involving
energies of the same order and nature as the electromagnetic radiation of light, have helped reveal that protein interactions are based on resonant electromagnetic energy transfer within the range of infra-red and visible light \cite{stoy3}. Formally one can describe the interaction of excitons with light by considering the model Hamiltonian studied in ref. \cite{stoy1}, with an extra term accounting for the exciton-photon coupling and the Maxwell equation describing the propagation of light. This results in the model studied recently \cite{stoy4,stoy5}, in which the authors established the existence of both bright and dark solitons associated with large-amplitude exciton and polariton excitations in the system. \par
However, the bright and dark soliton solutions proposed in refs. \cite{stoy4,stoy5} are localized structures with a vanishing shape at the boundaries of an infinitely long molecular chain. This is a shortcoming when considering their application in contexts where the molecular chain is finite, requiring periodic boundary conditions. Also, in real molecular crystals, dispersive interactions between molecular units are not limited only to nearest-neighbour molecules, the account of long-range interactions is relevant for a proper estimate not only of dispersion properties of excitations, but also characteristic parameters of bright and dark soliton structures. \par
In the present work we refomulate the generation and propagation of large-amplitude excitations in the model of ref. \cite{stoy4,stoy5}, taking into consideration the influence of long-range dispersive interactions between molecules along the molecular crystal. We focus on periodic soliton trains, instead of localized solitary-wave structures, for the first ones are more appropriate for molecular crystals with finite lengths. Proceeding with we consider the model Hamiltonian studied in ref. \cite{stoy5} which describes a longitudinal chain of interacting excitons coupled to electromagnetic radiations, but now taking into account the contribution of long-range intermolecular interactions on exciton dispersions. With this improved Hamiltonain, we obtain large-amplitude solutions to the coupled nonlinear exciton-polariton equations in terms of periodic short soliton wavetrains.  
\section{Model Hamiltonian and equations of motion}
\label{sec:one}
Consider a $1D$ discrete chain of molecules, in which Frenkel excitons have formed and interact with a propagating electromagnetic field. The total Hamiltonian for such system can be expressed \cite{stoy5}:
\begin{eqnarray}
H &=& \hbar\omega_0\sum_{n=1}^N{b^{\dagger}_n b_n} - \sum_{n=1}^N {\sum_{m\neq n}{J_{n-m}\left(b^{\dagger}_n b_m + b_n b^{\dagger}_m\right)}} \nonumber \\
 &-& D\sum_{n=1}^N{b^{\dagger}_n b_n b^{\dagger}_{n+1} b_{n+1}} - d \sum_{n=1}^N{\left(b_n^{\dagger}e_n^+ + b_ne_n^-\right)}, \label{eqa}
 \end{eqnarray}
 where $\hbar\omega_0$ is the energy associated with intramolecular vibrations, $J_{n-m}$ is the energy transfer integral between molecular excitations on sites $n$ and $m$, and $D$ is the exciton-exciton coupling strength. The Pauli operator $b^{\dagger}_n$ ($b_n$) creates (annihilates) a molecular exciton at site $n$, and we shall consider only dimer terms formed by nearest-neighbour excitons. The quantities $e^+$ and $e^-$ in eq. (\ref{eqa}) are the right-going and left-going components of the electromagnetic field, with $d$ the exciton-field coupling in the dipole approximation.  
 \par
 When $m=n\pm 1$, formula (\ref{eqa}) reduces to the Hamiltonian for a chain of excitons with only nearest-neighbour interactions \cite{stoy5}. We are interested in the system dynamics when long-range interactions between molecular units are fully accounted for. In this purpose, we assume a long-range interaction whose coupling strength falls off with a power law as one moves away from the first-neighbour molecules. One such long-range interaction is the so-called  Kac-Baker potential \cite{a10,a11,a12,a13}:
 \begin{equation}
  J_{n-m}= J_0 \frac{1-r}{2r}\, r^{\vert \ell\vert}, \hskip 0.2truecm \ell=m-n, \label{eqf}
 \end{equation}
in which the real parameter $r$ (with $0<r\leq 1$) determines the strength of intermolecular interactions with $\ell$ the distance between a molecule at site $n$ and a molecule at site $m$. For the long-range interaction potential eq. (\ref{eqf}), the first-neighbour model considered in ref. \cite{stoy5} is recovered when $r\rightarrow 0$. When $\ell \rightarrow \infty$ (or $r\rightarrow 1$), the intermolecular coupling becomes fully long ranged. As shown in ref. \cite{a14}, in this limit dispersive intermolecular interactions are purely of the “van der Waals” type. \par
According to its form, the long-range potential eq. (\ref{eqf}) applies only to physical systems for which $\ell=1$ or $\ell=\infty$ \cite{a11,a14}. In most real physical contexts of molecular crystals, however, long-range interactions are effectively sensitive only over a finite range of intermolecular distances, saturating beyond. Moreover, for molecular chains with finite lengths as it is the case for most biological systems, the maximum value of $\ell$ should be finite such that the Kac-Baker potential becomes unappropriate. As an alternative we shall consider the modified Kac-Baker potential \cite{a13}:
\begin{equation}
  J^L_{n-m}= \frac{J_0}{1-r^L} \frac{1-r}{2r}\, r^{\vert \ell\vert}, \hskip 0.2truecm \ell=m-n, \label{eqf1}
 \end{equation}
where $L$ is the maximum range (i.e. the maximum of $\ell$) beyond which intermolecular interactions saturate. In the nearest-neighbour limit (i.e. $\ell=L=1$), the quantity $J^L_{n-m}$ given in (\ref{eqf1}) reduces to $J_1= J_0/2$ regardless of the value of $r$. In the same limit, the Kac-Baker potential (\ref{eqf}) becomes $J_1= J_0(1-r)/2$. So for the Kac-Baker potential, only when $r=0$ one can recover the nearest-neighbour value of the intermolecular interaction strength. \par
\begin{figure*}
\centering
\begin{minipage}[c]{0.5\textwidth}
\includegraphics[width=3.0in, height=2.0in]{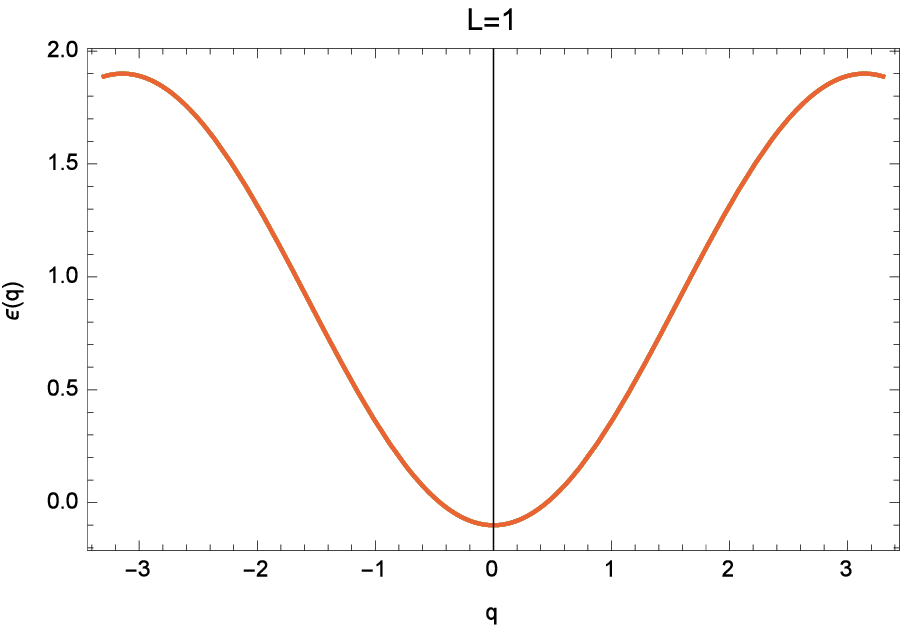}
\end{minipage}%
\begin{minipage}[c]{0.5\textwidth}
\includegraphics[width=3.0in, height=2.0in]{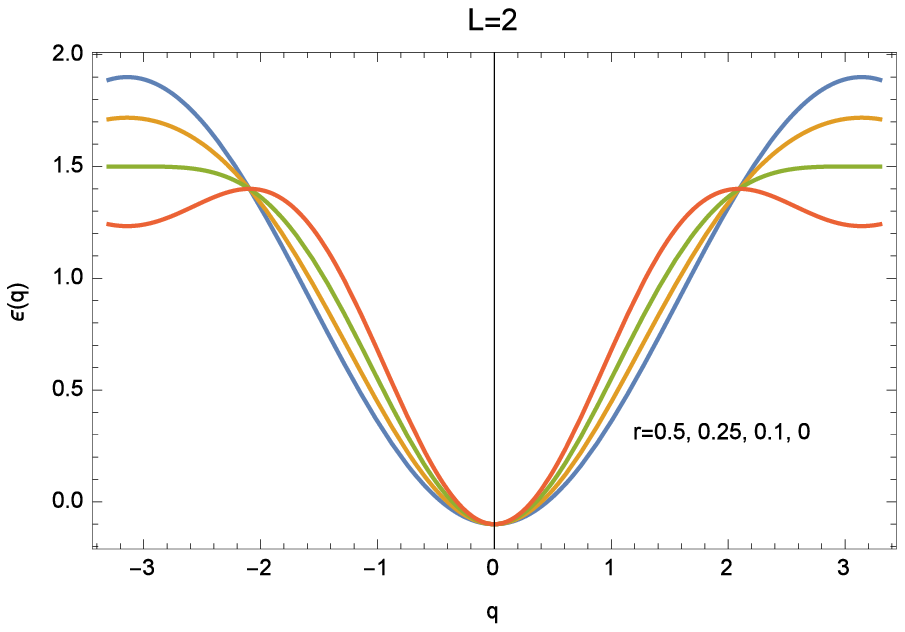}
\end{minipage}\\
\begin{minipage}[c]{0.5\textwidth}
\includegraphics[width=3.0in, height=2.0in]{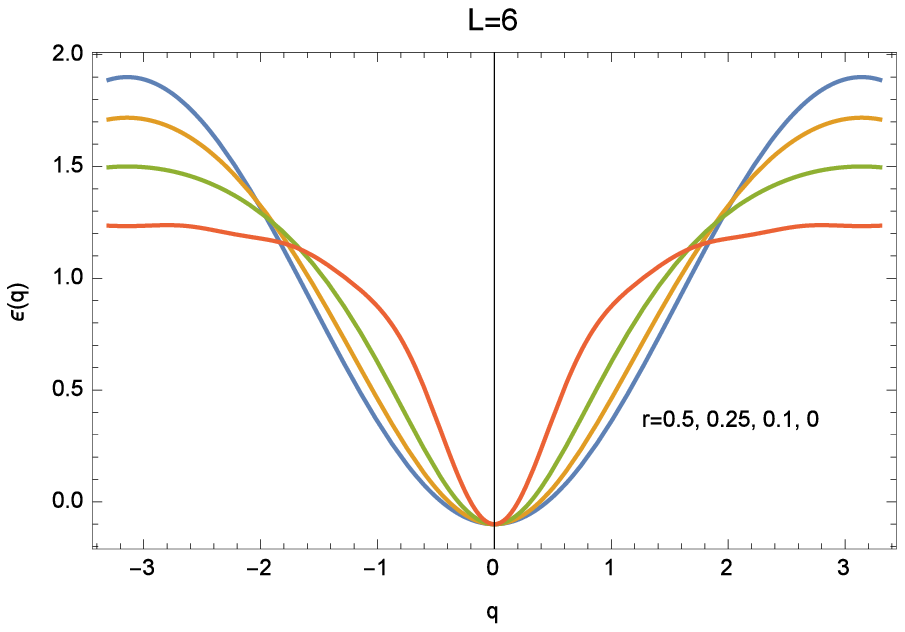}
\end{minipage}%
\begin{minipage}[c]{0.5\textwidth}
\includegraphics[width=3.0in, height=2.0in]{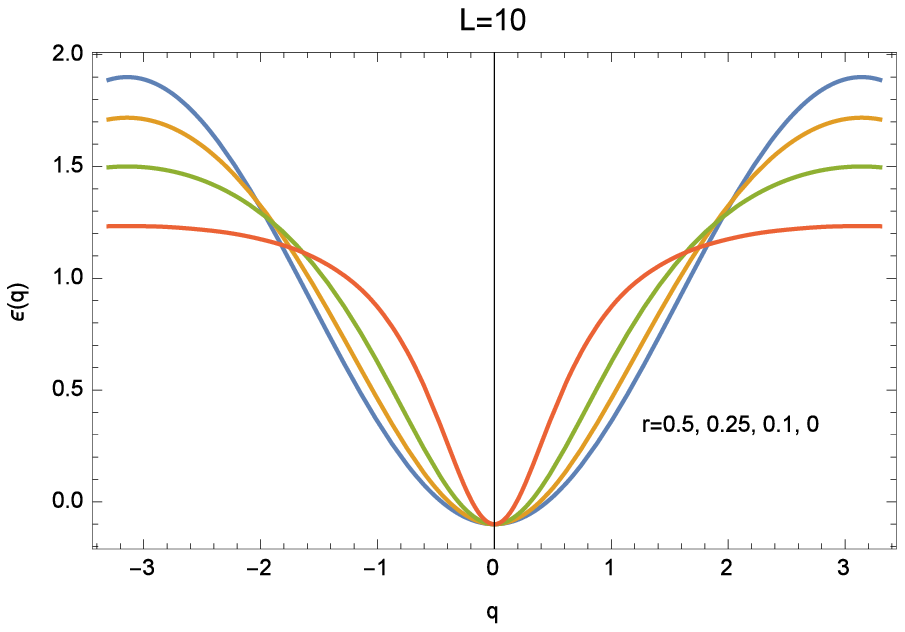}
\end{minipage}
\caption{\label{fig:one} (Color online) Single-exciton energy versus wavenumber $q$ (in units of the lattice spacing $a$), for different values of $r$ and $L$.}
\end{figure*}
\begin{figure*}
\centering
\begin{minipage}[c]{0.5\textwidth}
\includegraphics[width=3.0in, height=2.0in]{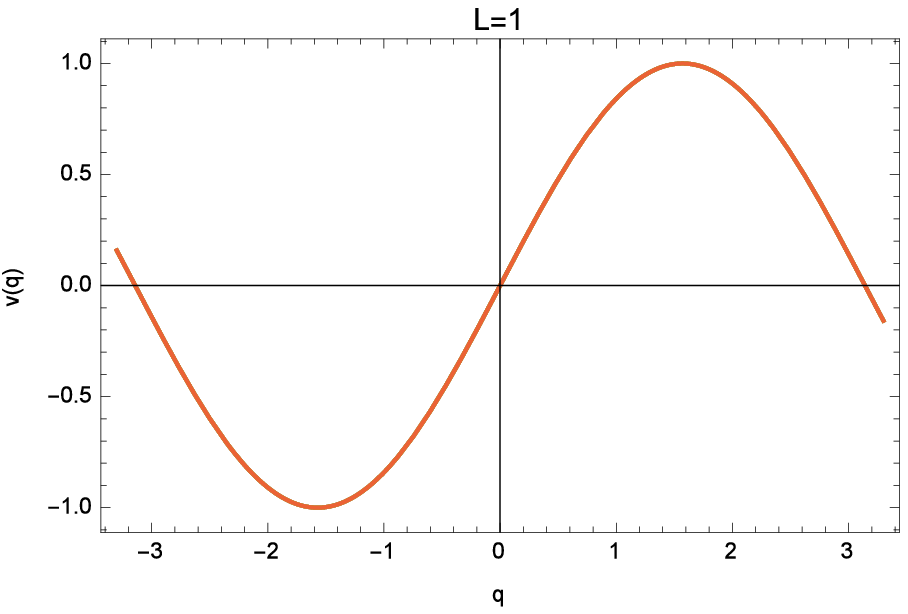}
\end{minipage}%
\begin{minipage}[c]{0.5\textwidth}
\includegraphics[width=3.0in, height=2.0in]{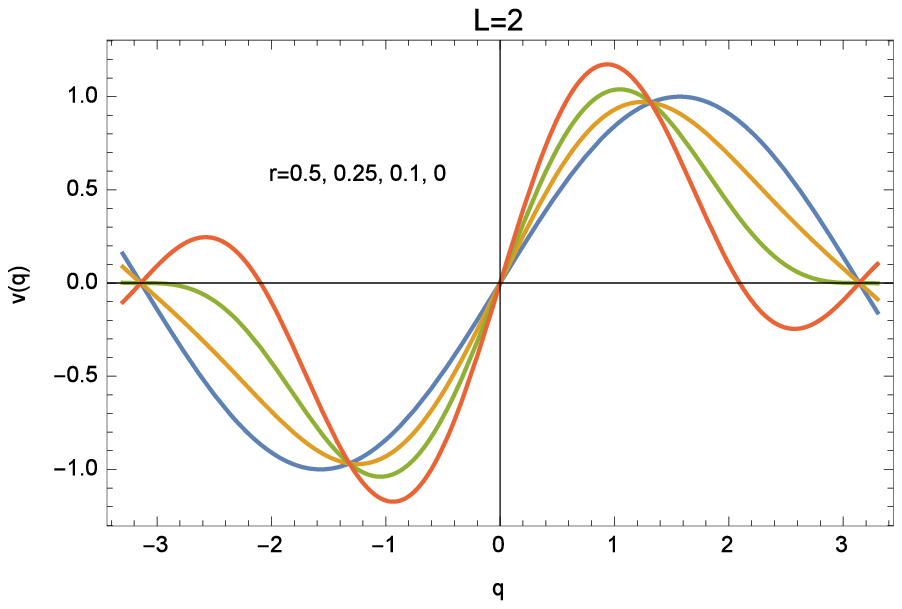}
\end{minipage}\\
\begin{minipage}[c]{0.5\textwidth}
\includegraphics[width=3.0in, height=2.0in]{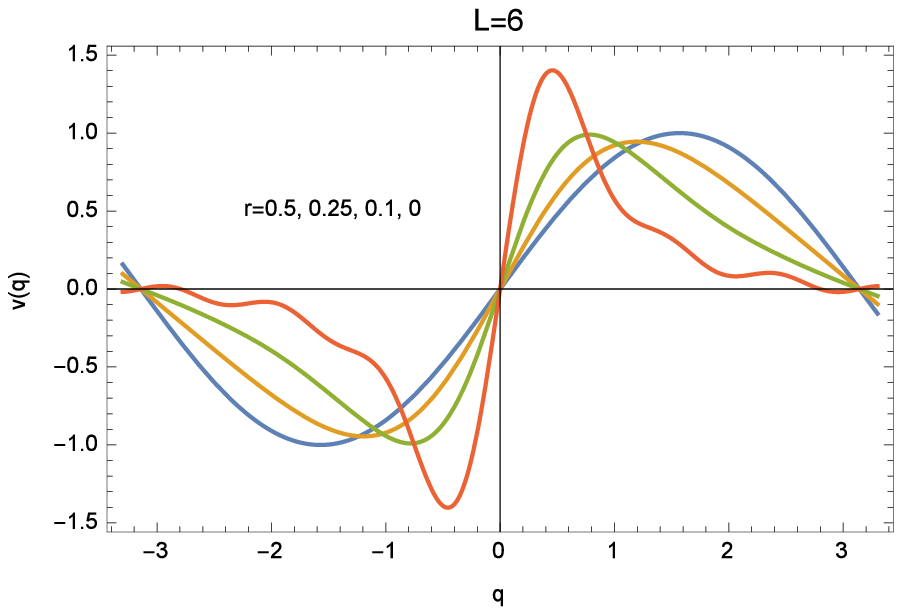}
\end{minipage}%
\begin{minipage}[c]{0.5\textwidth}
\includegraphics[width=3.0in, height=2.0in]{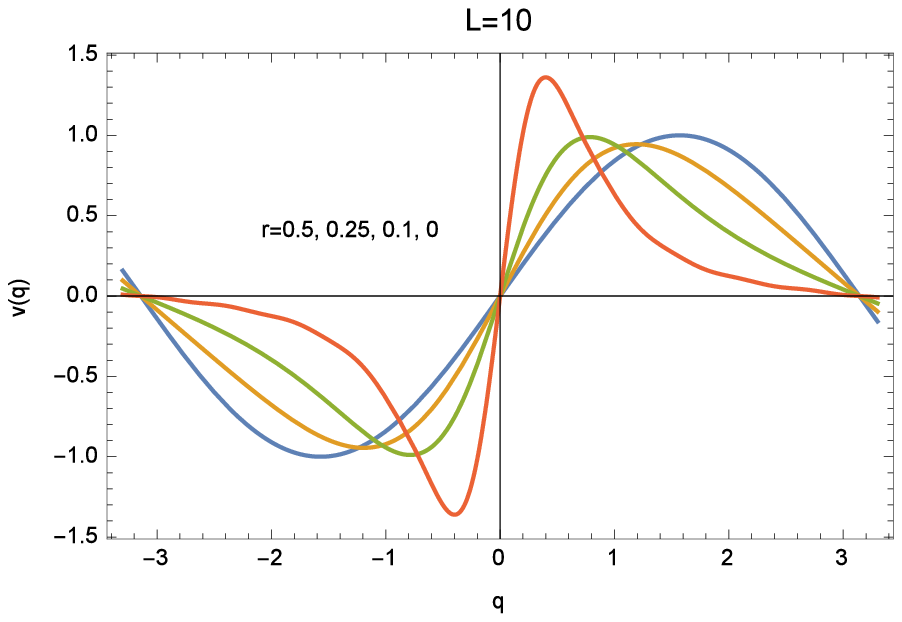}
\end{minipage}
\caption{\label{fig:two} (Color online) Single-exciton velocity versus wavenumber $q$ (in units of the lattice spacing $a$), for different values of $r$ and $L$.}
\end{figure*}

To derive the equation of motion for excitons from the Heisenberg formalism on the Hamiltonian formula (\ref{eqa}), it is useful to start by remarking that being Pauli operators the quantities $b_n$ and $b_n^ {\dagger}$ obey the commutation relations: 
\begin{equation}
\left[b_n, b_m^{\dagger}\right]=(1-2P_n)\delta_{n,m}, \hskip 0.25truecm P_n=b_n^{\dagger}b_n.
\end{equation}
With this, the Heisenberg formalism leads to the following exciton equations of motion:
 \begin{eqnarray}
i\hbar \frac{\partial b_n}{\partial t}&=& \hbar\omega_0 b_n - (1-2P_n)\sum_{m\neq n}{J^L_{n-m} b_m} \nonumber \\
 &-& 2Db_n \left(b^{\dagger}_{n+1} b_{n+1} + b^{\dagger}_{n-1} b_{n-1}\right) \nonumber \\
 &-& d(1-2P_n)e_n^+. \label{eqb}
 \end{eqnarray}
Let us seek for coherent structures associated with the dynamics of single-particle states, in this goal we can readily approximate $\langle P_n\rangle\approx \langle b_n\rangle \langle b_n^{\dagger}\rangle$. Setting $\langle b_n\rangle=\alpha_n$ where $\alpha_n$ is the single-exciton occupation probability \cite{dak1}, eq. (\ref{eqb}) becomes: 
\begin{eqnarray}
i\hbar \frac{\partial \alpha_n}{\partial t}&=& \hbar\omega_0 \alpha_n - (1-2\vert \alpha_n\vert^2)\sum_{m\neq n}{J^L_{n-m} \alpha_m} \nonumber \\
 &-& 2D\alpha_n \left(\vert \alpha_{n+1}\vert^2 + \vert \alpha_{n-1}\vert^2\right) \nonumber \\
 &-& d(1-2\vert\alpha_n\vert^2)e_n^+. \label{eqba}
 \end{eqnarray}
As for the electromagnetic field $e_n$, we assume the optical response of the medium to the field propagation is linear. In this case the field propagation can be described by the linear Maxwell equation with a "source term" i.e.:
\begin{equation}
\left(\frac{\partial^2}{\partial x^2} - \frac{1}{c^2}\frac{\partial^2}{\partial t^2}\right)e^+(x, t)= \frac{4\pi d}{c^2a^3}\frac{\partial^2}{\partial t^2}\alpha(x,t), \label{eqmax}
\end{equation}
where $c$ is light speed in the molecular crystal and $a$ is the lattice spacing. Eq. (\ref{eqmax}) suggests that the field propagates in a continuous medium, therefore we must carry out a continuum-limit approximation on the exciton variable $\alpha_n$. To this end we define the continuous position $x=na$, such that $\alpha_{n\pm 1}\approx \alpha(x, t)\pm a\frac{\partial}{\partial x}\alpha(x,t) + \frac{a^2}{2} \frac{\partial^2 }{\partial x^2}\alpha(x,t) + ...$. Treating similarly the term containing the long-range interaction potential $J^L_{m-n}$ \cite{edis}, and seeking for solutions of the forms:
\begin{equation}
\alpha(x,t)=e^{i(qx-\omega t)}\phi(x,t), \hskip 0.3truecm e^+(x,t)= e^{i(qx-\omega t)}\varepsilon(x,t), \label{newsol}
\end{equation}
where $\phi(x,t)$ and $\varepsilon(x,t)$ are now purely reals, and $\omega$ and $q$ and the exciton frequency and wavenumber respectively, eqs. (\ref{eqba}) and (\ref{eqmax}) reduce to:
\begin{eqnarray}
i\hbar \frac{\partial \phi(x,t)}{\partial t}&=& (\epsilon_L(q) - \omega)\phi(x,t) - i\hbar v_L(q) \frac{\partial \phi(x,t)}{\partial x} \nonumber \\
&-& D_L(q)\frac{\partial^2 \phi(x,t)}{\partial x^2} -\lambda_L(q) \phi^3 \nonumber \\
&-& d\lbrack 1-2\phi(x,t)^2\rbrack \varepsilon(x,t), \label{qe1}
\end{eqnarray}
\begin{eqnarray}
[c^2(\frac{\partial^2}{\partial x^2} &-& \frac{1}{c^2}\frac{\partial^2}{\partial t^2}) + 2i(c^2 q\frac{\partial}{\partial x} + \omega\frac{\partial}{\partial t}) \nonumber \\
&+& (\omega^2-c^2q^2)] \varepsilon(x,t) \nonumber \\
  &=& \frac{4\pi d}{a^3}\lbrack\frac{\partial^2}{\partial t^2} - \omega^2-2i\omega\frac{\partial }{\partial t}\rbrack \phi(x,t), \label{eqmaxa}
\end{eqnarray}
with:
\begin{eqnarray}
\epsilon_L(q)&=&\hbar\omega_0 - 2\sum_{\ell=1}^L {J_{\ell} cos(\ell q a)}, \nonumber \\
\lambda_L(q)&=& 4D - 2(\hbar\omega_0 - \epsilon_L(q)), \label{disa} \\
v_L(q)&=& \frac{1}{\hbar}\frac{d\epsilon_L(q)}{dq}, \nonumber \\
D_L(q)&=& \hbar\frac{dv_L(q)}{dq}=\frac{d^2\epsilon_L(q)}{d q^2}. \label{disp}
\end{eqnarray}
The sum over $\ell$ in the expression of $\epsilon_L(q)$ in eq. (\ref{disa}) is exact \cite{a13}, yielding:
\begin{eqnarray}
\epsilon_L(q)&=&\hbar\omega_0 - J_0 \frac{1-r}{1-r^L} \frac{\cos(qa) - r-r^L K_L(q)}{1-2r\cos(qa)+r^2}, \label{pta}  \\
K_L(q)&=&\cos \left[(L+1)qa\right] - r\cos(qLa). \label{pt}
\end{eqnarray}
Introducing the new coordinate $z=x-vt$ and applying the slowly-varying envelope approximation \cite{stoy4,stoy5}, eqs. (\ref{qe1})-(\ref{eqmaxa}) now read:
\begin{equation}
iP_L(q)\frac{\partial \phi}{\partial z} + M_L(q)\frac{\partial^2 \phi}{\partial z^2} - \chi_L(q)\phi+ A_L(q)\phi^3=0, \label{asw1} 
\end{equation}
\begin{equation}
\frac{\partial^2 \varepsilon}{\partial z^2}=\frac{1}{d}(\epsilon_L(q)-\hbar\omega)\frac{\partial^2\phi}{\partial z^2}, \label{asw}
\end{equation}
where:
\begin{eqnarray}
P_L(q)&=&\hbar(v_L(q)-v) \nonumber \\
&+& \frac{1}{c^2q^2 - \omega^2}\lbrack 2(c^2q-\omega v)(\epsilon_L(q)-\hbar\omega)-2\hbar\omega v\Omega_0\rbrack, \nonumber \\
\Omega_0&=& \frac{4\pi d^2}{\hbar a^3}, \hskip 0.25truecm \chi_L(q)=\epsilon_L(q)-\hbar\omega - \frac{\hbar\Omega_0 \omega^2}{c^2q^2-\omega^2}, \nonumber \\
A_L(q)&=& \lambda_L(q)- \frac{\hbar\Omega_0 \omega^2}{c^2q^2-\omega^2}, \nonumber \\
M_L(q)&=&D_L(q)+\frac{1}{c^2q^2 - \omega^2}[2\hbar(c^2q-\omega v)(v_L(q)-v) \nonumber \\
&+& (c^2-v^2)(\epsilon_L(q)-\hbar\omega)-\hbar\Omega_0 v^2]. \nonumber \\
\end{eqnarray}
In general, real solutions to the nonlinear equation (\ref{asw1}) is possible provided $P_L(q)=0$. This condition implies that the solution is a nonlinear wave travelling at velocity: 
\begin{equation}
v=\frac{2c^2q(\epsilon_L(q)-\hbar\omega)+(c^2q^2 - \omega^2)\hbar v_L(q)}{2\omega(\epsilon_L(q)-\hbar\omega + \hbar\Omega_0) +\hbar(c^2q^2-\omega^2)}. \label{vitesa}
\end{equation}
It turns out that because the energy $\epsilon_L(q)$ and velocity $v_L(q)$ of single-exciton states depend on the intermolecular interaction strength $r$ and range $L$, the velocity $v$ of nonlinear excitations governed by eq. (\ref{asw1}) will be sensitive to long-range intermolecular interactions. To privide a sight on the effects of long-range intermolecular interactions on these two relevant parameters, in figs. \ref{fig:one} and \ref{fig:two} we plotted $\epsilon_L(q)$ and $v_L(q)$ respectively as a function of $q$, for different values of the intermolecular interaction strength $r$ and range $L$. For all the curves we used the following (arbitrary) values for the model parameters: $J_0 = 1$, $\omega_0 = 0.9$, $\omega = 1.2$, $\Omega_0= 0.1$, $a=1$.

Fig. \ref{fig:one} shows that as we increase the interaction range $L$, the upper edge of the Brillouin zone is shifted from its short-range value at $q=\pi$. This shift causes a shrinkage of the Brillouin zone as $L$ increases, and a decrease of the long-wavelength cut-off (i.e. low-lying single-exciton) energy $\epsilon_L(q=0)$ as $r$ is increased. It is also remarkable that the slope of the single-exciton dispersion energy $\epsilon_L(q)$, near $q= 0$, is increased as the strength of the long-range interaction $r$ increases. This actually indicates an increase of the characteristic speed of low-lying excitons, as the range and strength of intermolecular interactions are simultaneously increased. This last behaviour is more evident in fig. \ref{fig:two}, representing the velocity dispersion of single-exciton modes $q$ for different values of $r$ and $L$. \par
In the next section we derive the two possible solutions to the nonlinear eq. (\ref{asw1}), laying emphasis on periodic wavetrains formed of bright and dark solitons of equal finite separation (i.e. finite repetition). Shape profiles of the electromagnetic wave coupled to the exciton, for each type of periodic soliton solution obtained, will follow from eq. (\ref{asw}). 
\section{Periodic soliton train solutions}
\label{sec:two}
\subsection{Periodic trains of bright solitons}
The single-soliton solution to the nonlinear equation (\ref{asw1}) depends on signs of $M_L(q)$ and $A_L(q)$. For single-pulse (i.e. bright) soliton solution propagating with velocity $v$ given by (\ref{vitesa}), the two parameters must be of the same sign. Applying the boundary conditions $\phi(z\rightarrow \pm)\rightarrow 0$, we find the localized pulse solution:
\begin{equation}
\phi(x,t)= \phi_0\,sech\frac{x-vt}{\Gamma}, \label{br}
\end{equation}
\begin{equation}
\Gamma= \sqrt{\frac{2M_L(q)}{\chi_L(q)}}, \hskip 0.25truecm \phi_0= \sqrt{\frac{2\chi_L(q)}{A_L(q)}}. \label{ta1}
\end{equation} 
In our case we are interested in a solution describing a periodic train of pulses of the form (\ref{br}), forming a pulse crystal with a finite period. We can therefore express such solution formally as \cite{dikaa}:
\begin{equation}
\phi_{\nu}(z)=\sum_{i=0}^Q{\phi(z-i\nu)}, \label{sola}
\end{equation}
where $\phi(z)$ is the single-pulse solution (\ref{br}). Eq. (\ref{sola}) represents a lattice of $Q$ identical pulses of equal separation $\nu$, when $Q\rightarrow \infty$ the sum becomes exact and we find \cite{dikaa,dikaro}:
\begin{equation}
\phi_{\nu}(x,t)=\frac{\phi_0}{\sqrt{2-\kappa^2}}\,dn\frac{x-vt}{\Gamma_{\kappa}}, \hskip 0.25truecm \Gamma_{\kappa}=\sqrt{2-\kappa^2}\Gamma, \label{dnsol}
\end{equation}
where $dn()$ is a Jacobi elliptic function \cite{dikaa,dikaro,dikab,dikac,dikad}. The solution (\ref{dnsol}) can also be obtained directly by solving eq. (\ref{asw1}) with boundary conditions $\phi(z\pm \nu)= \phi(z)$, yielding the well-known \cite{dikaa,dikaro,dikab,dikac,dikad} periodic soliton solution to the nonlinear Schr\"odinger equation. More explicitely, formula (\ref{dnsol}) describes a periodic lattice of bright solitons of identical amplitude $\phi_0/\sqrt{2-\kappa^2}$, identical average width $\Gamma_{\kappa}$ and period:
\begin{equation}
\nu= 2K(\kappa)\Gamma_{\kappa},
\end{equation}
with $K(\kappa)$ the elliptic integral of first kind and $\kappa$ ($0<\kappa \leq 1$) the modulus of the Jacobi elliptic function. It is quite remarkable that when $\kappa\rightarrow 1$, the function $dn()\rightarrow sin()$ and when $\kappa\rightarrow 1$ the function $dn()\rightarrow sech()$. In this last limit the period $\nu\rightarrow \infty$ and the soliton amplitude reduces to $\phi_0$, while its average width becomes $\Gamma$. 
\subsection{Periodic trains of dark solitons}
When $M_L(q)$ and $A_L(q)$ are of different signs, the nonlinear wave equation (\ref{asw1}) with boundary conditions $\phi(z\rightarrow\pm \infty)=\pm \phi_1$ admits a single-kink soliton solution: 
\begin{equation}
\phi_1(x,t)= \phi_1\,tanh\frac{x-vt}{\Gamma_1}, \label{br1}
\end{equation}
\begin{equation}
\Gamma_1= \sqrt{\frac{-2M_L(q)}{\chi_L(q)}}, \hskip 0.25truecm \phi_1= \sqrt{\frac{\chi_L(q)}{A_L(q)}}. \label{ta2}
\end{equation} 
Formula (\ref{ta2}) suggests that $\chi_L(q)$ and $A_L(q)$ should be positive, while $M_L(q)$ should be negative for a stable single-kink soliton solution. The wavetrain structure associated with the single-kink soliton (\ref{br1}), consisting of a periodic arrangement of identical kinks of the form (\ref{br1}) with equal separation $\nu_1$, is obtained either by summing $\phi_1(z-i\nu)$ over $i$ from zero to $\infty$, or solving directly eq. (\ref{asw1}) with periodic boundary conditions. The two approaches result in the following nonlinear periodic solution:
\begin{equation}
\phi_{\nu_1}(x,t)=\sqrt{\frac{2\kappa^2}{1+\kappa^2}}\phi_1\,sn\frac{x-vt}{\Gamma_{1,\kappa}}, \hskip 0.25truecm \Gamma_{1,\kappa}=\sqrt{\frac{1+\kappa^2}{2}}\Gamma_1, \label{dnsol1}
\end{equation}
where $sn()$ is another Jacobi elliptic function with a period \cite{dikaq,dikaq1}:
\begin{equation}
\nu_1= K(\kappa)\Gamma_{1,\kappa}.
\end{equation} 
As in the case of the periodic bright soliton lattice (\ref{dnsol}), the periodic kink solution (\ref{dnsol1}) tends to an harmonic wave when $\kappa=0$ and reduces exactly to the single-kink solution (\ref{br1}) when $\kappa=1$. Note that in this last limit the period $\nu_1\rightarrow \infty$. \par
\begin{figure*}
\centering
\begin{minipage}[c]{0.5\textwidth}
\includegraphics[width=3.0in, height=2.0in]{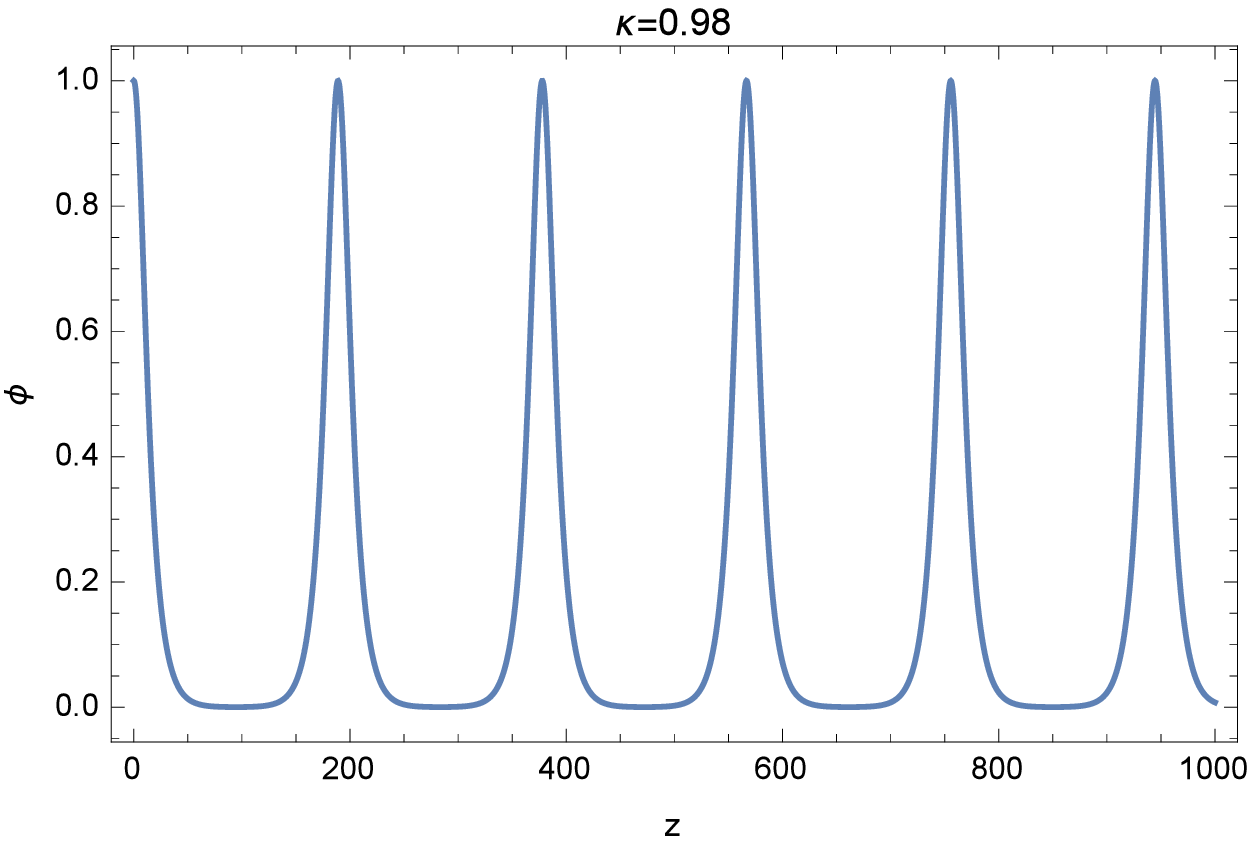}
\end{minipage}%
\begin{minipage}[c]{0.5\textwidth}
\includegraphics[width=3.0in, height=2.0in]{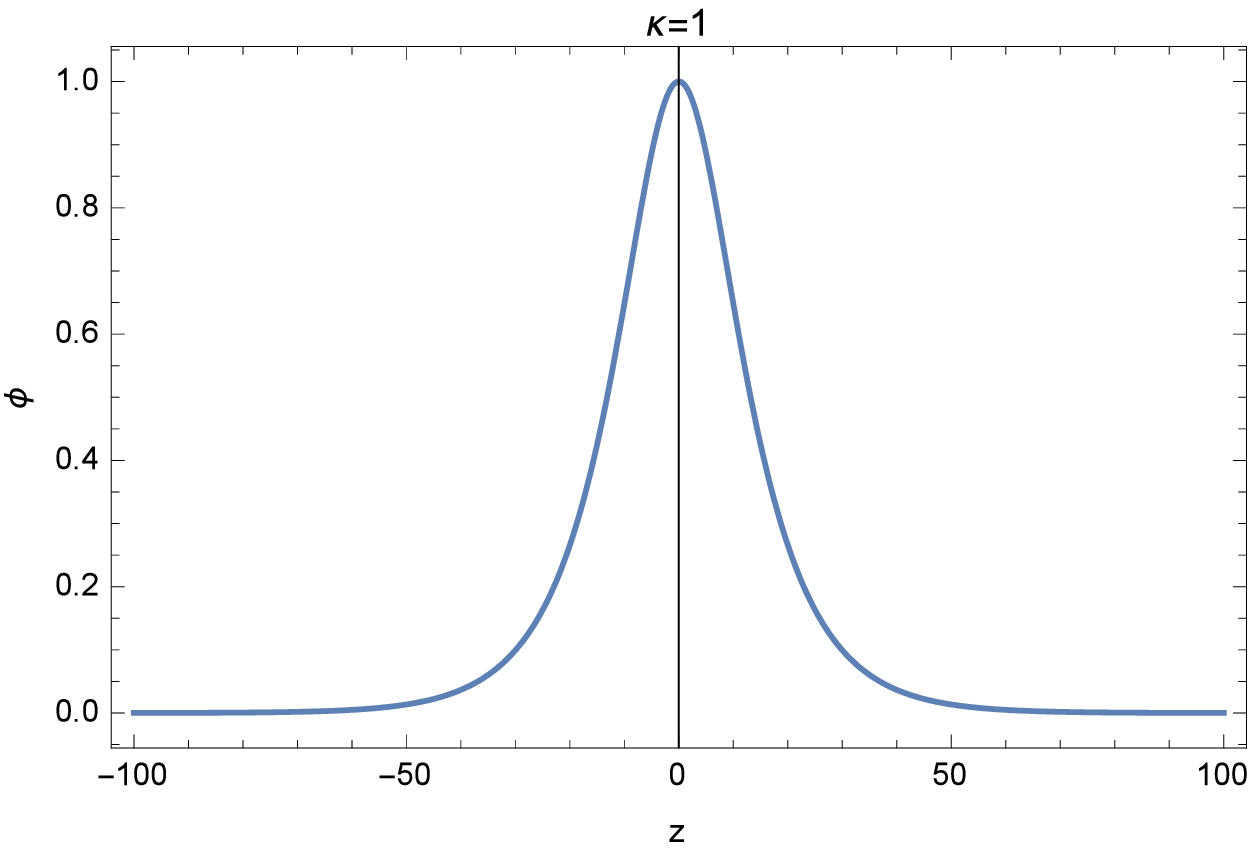}
\end{minipage}\\
\begin{minipage}[c]{0.5\textwidth}
\includegraphics[width=3.0in, height=2.0in]{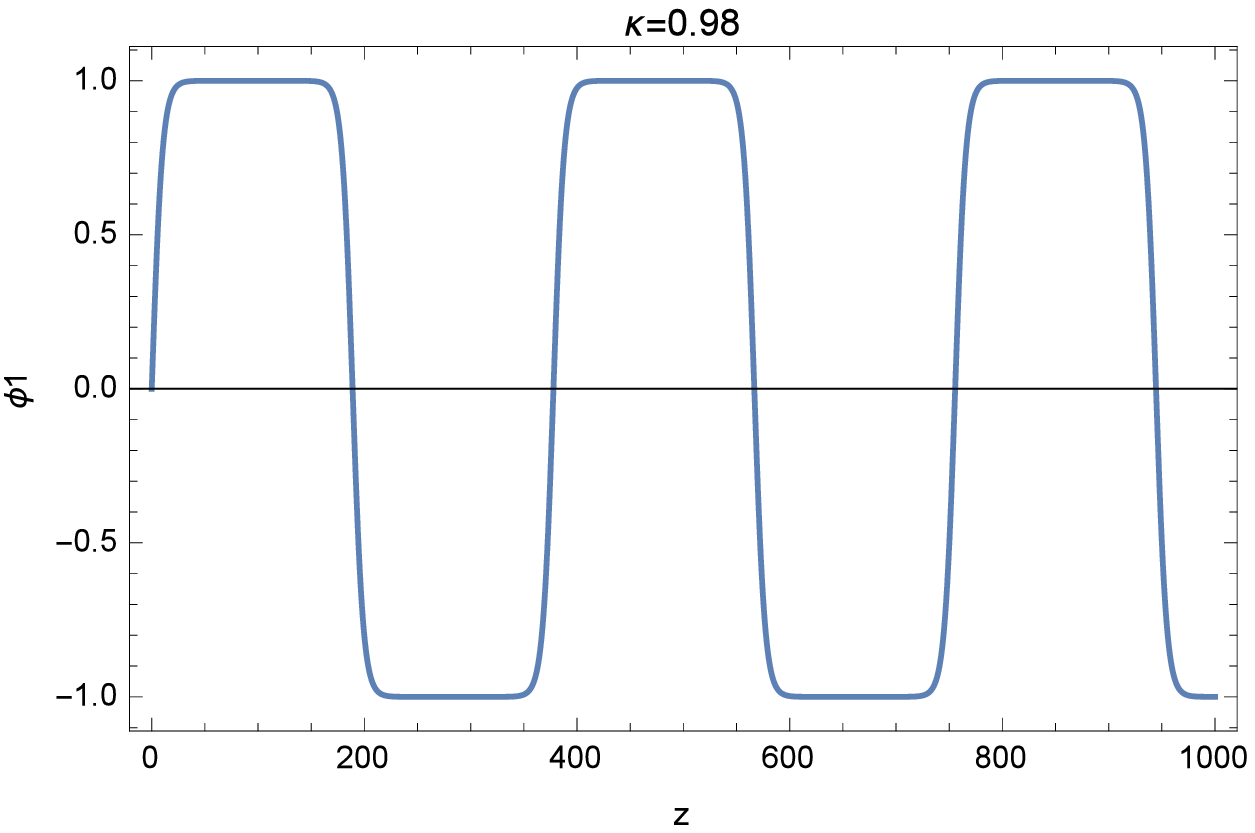}
\end{minipage}%
\begin{minipage}[c]{0.5\textwidth}
\includegraphics[width=3.0in, height=2.0in]{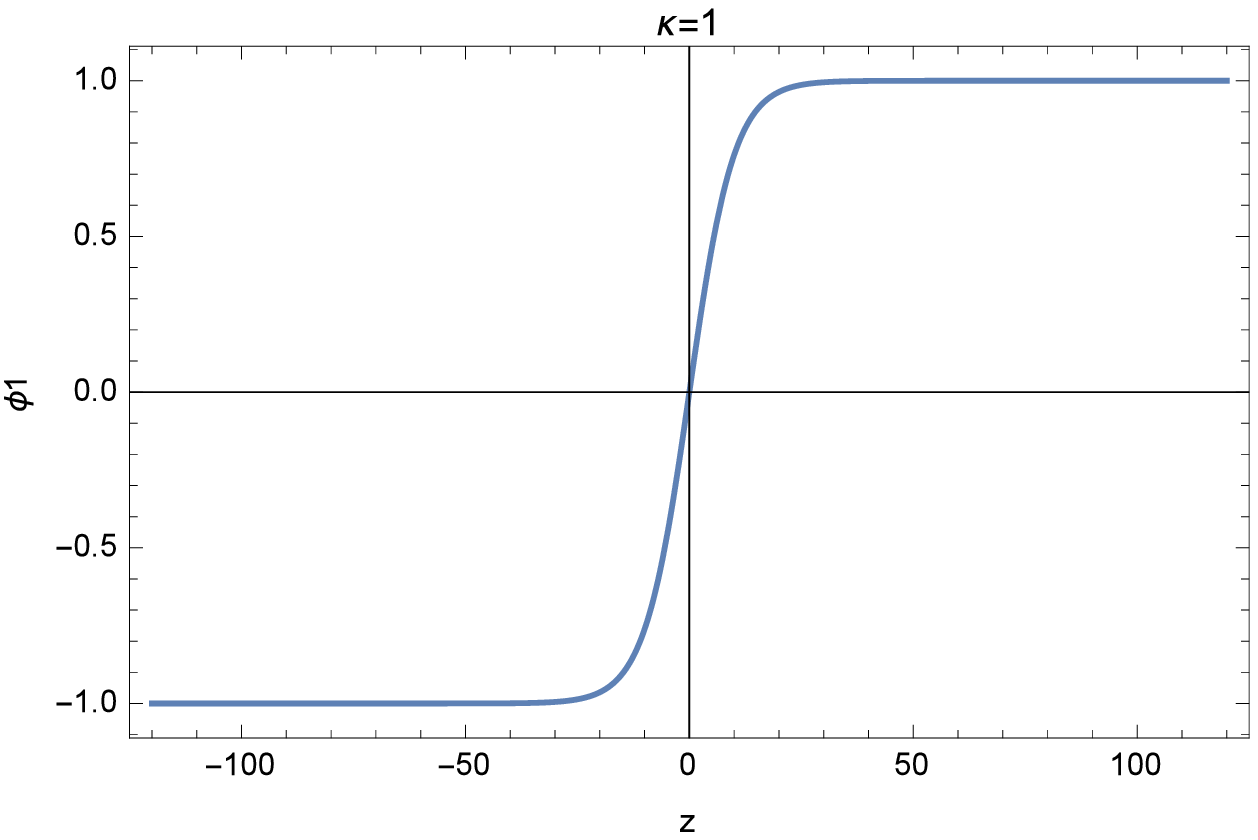}
\end{minipage}
\caption{\label{fig:three} (Color online) Sketches of the bright (upper graphs) and dark (lower graphs) soliton solutions obtained in (\ref{dnsol}) and (\ref{dnsol1}) respectively, for two values of $\kappa$. Left graphs are for $\kappa=0.98$, while the right graphs are the single-soliton limit (i.e. $\kappa=1$).}
\end{figure*}

Formula (\ref{ta1}) and (\ref{ta2}) suggest that in addition to their dependence on $\kappa$, the amplitudes and widths of the single-pulse and single-kink solitons composing the bright- and dark-soliton trains eq. (\ref{dnsol}) and (\ref{dnsol1}), are functions of the long-range parameters $r$ and $L$. This dependence is introduced essentially through the energy $\epsilon_L(q)$ and velocity $v_L(q)$ of the single-exciton states, which variations with characteristic parameters of the long-range intermolecular interaction potential were illustrated in figs. \ref{fig:one} and \ref{fig:two}. In fig. \ref{fig:three} we sketched the bright (upper graphs) and dark (lower graphs) soliton wavetrains obtained in formula (\ref{dnsol}) and (\ref{dnsol1}), for $\kappa=0.98$ and $\kappa=1$. The left graphs are the single-soliton limits which, as indicated, are recovered when $\kappa=1$ (i.e. when the separation between the single-soliton components of the two soliton wavetrains is infinitely large).

\section{Concluding remarks}
\label{sec:three}
Molecular crystals offer a wealth of rich physical properties that are distinct from those observed in conventional solids such as covalent or ionic crystals. In molecular solids, the packing of molecules to form the molecular crystal structure is determined to a large extent by relatively weaker intermolecular interactions such as hydrogen bonds and van der Waals. Organic molecular crystals \cite{org} in particular are interesting from both biological and biochemical perspectives, they provide ideal candidates for unveiling the nature of the hydrogen bond and hence to understand a variety of important biological processes in molecular solids such as proteins and peptides. \par
In this work we investigated the effects of long-range intermolecular interactions on the dispersion energy of single-exciton states, and by extension on characteristic parameters of periodic wavetrains of exciton and polariton solitons in molecular crystals interacting with light. By reformulating the usual Kac-Baker potential to take into account the finite range of intermolecular interactions, we found that the account of long-range interactions brings a huge quantitative as well as qualitative changes on exciton and polariton solitons characteristic parameters. Namely we obtained that long-range interactions will narrow the Brillouin zone of small-amplitude excitons with respect to the short-range result, while enhancing their acoustic velocity but lowering the single-exciton low-lying energy state. Large-amplitude solutions to the exciton equation was obtained in the continuum limit in terms of periodic wavetrains of pulse and dark solitons, whose amplitudes and widths as well as periods were affected by the long-range intermolecular interactions. single and periodic soliton solutions to the light propagating coupled to the excitons, are similar to the exciton solitons for they are linked by a simple linear relation. \par
Molecular crystals today are used for a variety of technological applications, in these applications the crystal structure determined by the intermolecular interactions such as hydrogen bonds and van der Waals interactions, is key in the manipulation of the structure so as to produce materials with technologically useful properties as for instance nonlinear optical responses, short exciton pulses, robust exciton transports and so on \cite{org}.

\section*{Acknowledgments}
Work supported by the World Academy of Sciences (TWAS), Trieste, Italy. A. M. Dikand\'e wishes to thank the Alexander von Humboldt (AvH) Foundation of logistic supports. 
\section*{Authors contribution statement}
ENNA performed calculations, AMD proposed the theme and wrote the manuscript. Both authors agreed on curves which were plotted by AMD.

\end{document}